\documentclass[twocolumn,showpacs,preprintnumbers,amsmath,amssymb,superscriptaddress]{revtex4}
\usepackage{graphicx}
\usepackage{dcolumn}
\usepackage{bm}
\newcommand{\mypath}[1]{./#1}

\begin{document}
\title{Superconducting Puddles and ``Colossal'' Effects \\
in Underdoped Cuprates}


\author{Gonzalo Alvarez}
\affiliation{National High Magnetic Field Lab and Department of Physics, 
Florida State University, Tallahassee, FL 32310}

\author{Matthias Mayr}
\affiliation{Max-Planck-Institut f\"ur Festk\"orperforschung, 70569 Stuttgart, Germany.}

\author{Adriana Moreo}
\affiliation{National High Magnetic Field Lab and Department of Physics, 
Florida State University, Tallahassee, FL 32310}

\author{Elbio Dagotto}
\affiliation{National High Magnetic Field Lab and Department of Physics, 
Florida State University, Tallahassee, FL 32310}

\date{\today}

\begin{abstract}
Phenomenological models for the antiferromagnetic (AF)
vs.\ $d$-wave superconductivity competition in cuprates
are studied using conventional Monte Carlo techniques.
The analysis suggests that cuprates 
may show a variety of different behaviors in the very underdoped regime:
local coexistence or first-order transitions among the competing orders,
stripes, or glassy states with nanoscale
superconducting (SC) 
puddles. The transition from AF to SC does not seem universal.
In particular, the glassy state leads to the possibility of ``colossal''
effects in some cuprates, analog of those in manganites.
Under suitable conditions, non-superconducting Cu-oxides
could rapidly become superconducting by the influence of weak 
perturbations that align the randomly oriented phases
of the SC puddles in the mixed state. Consequences of these ideas
for thin-film and photoemission experiments are discussed.
\end{abstract}

\maketitle

\section{Introduction}

Clarifying the physics of high-temperature superconductors (HTS) is still
one of the most important challenges in condensed-matter physics. 
There is overwhelming experimental 
evidence for several unconventional regimes in HTS, including 
a pseudogap region at temperatures above the superconducting (SC) phase, and
a largely unexplored glassy state separating the parent antiferromagnet (AF) 
from the SC phase at low hole-doping $x$.
In addition, recent investigations unveiled another
remarkable property of HTS's that defies conventional wisdom:
the existence of {\it giant proximity effects} (GPE) 
in some cuprates,\cite{re:bozovic04,re:decca00,re:quintanilla03}
where a supercurrent  in Josephson junctions was found to run through non-SC
Cu-oxide-based thick barriers. This contradicts
the expected exponential suppression of
supercurrents with barrier thickness beyond the short
coherence length of Cu-oxides. 
The purpose of this paper is to propose an explanation based
on a description of the glassy state as containing SC {\it puddles}.
This nanoscale inhomogeneous state leads 
to {\it colossal
effects in cuprates}, in analogy with 
manganites\cite{re:dagotto02,re:dagotto01,re:tokura00b}. 
In addition, it is argued that different
inhomogeneous states could be stabilized in different Cu-oxides,
depending on coupling and quenched disorder strengths.
In fact, neutron scattering studies have revealed
``stripes'' of charge in Nd-LSCO,\cite{re:tranquada95,re:emery90} but scanning
tunneling microscopy (STM) experiments\cite{re:lang02,re:kohsaka04} indicate ``patches'' 
in Bi2212, consistent 
with our analysis. There is no unique way to transition from AF to SC.

Studies of the t-J model have revealed SC
and striped states\cite{re:sorella02,re:white98} evolving from the undoped limit. Then, it is
reasonable to assume that AF, SC, and striped states are 
dominant in cuprates, and their competition regulates the
 HTS phenomenology. However, further computational 
progress using basic models is
limited by cluster sizes that
cannot handle the nanoscale structure unveiled by STM experiments.
Considering these restrictions, 
here a $phenomenological$ approach will
be pursued to understand how these phases compete, 
incorporating the quenched disorder inevitably introduced by
chemical doping. This effort unveils novel effects of experimental 
relevance, not captured with first-principles studies.
Two models are used, one with itinerant fermions and the other without,
and the conclusions are similar in both. Hopefully, this effort will jump start a more detailed computational 
analysis of
phenomenological models in the high-Tc arena, since most basic first-principles approaches, including Hubbard and $t-J$ investigations,
  have basically reached their
limits, particularly regarding cluster sizes that can be studied.

\section{Model I: Itinerant Fermions}
The analysis starts with a phenomenological model of itinerant
electrons (simulating carriers) 
on a square lattice, locally coupled to classical order
parameters:
\begin{eqnarray}\nonumber
H_{\rm F}&=&-t\sum_{<{\bf ij}>,\sigma}(c^\dagger_{{\bf i}\sigma}c_{{\bf j}\sigma}
+H.c.) + 2 \sum_{\bf i} J_{\bf i} S_{\bf i}^z s_{\bf i}^z
-\sum_{{\bf i}\sigma} \mu_{\bf i} n_{{\bf i}\sigma} \\ 
& + & 
\frac {1}{D}\sum_{{\bf i},\alpha}\frac {1}{V_{\bf i}}|\Delta_{{\bf i}\alpha}|^2-
\sum_{{\bf i},\alpha}(\Delta_{{\bf i}\alpha}
c_{{\bf i}\uparrow}c_{{\bf i}+\alpha\downarrow}+H.c.), 
\label{eq:hamfermi}
\end{eqnarray}

\begin{figure}
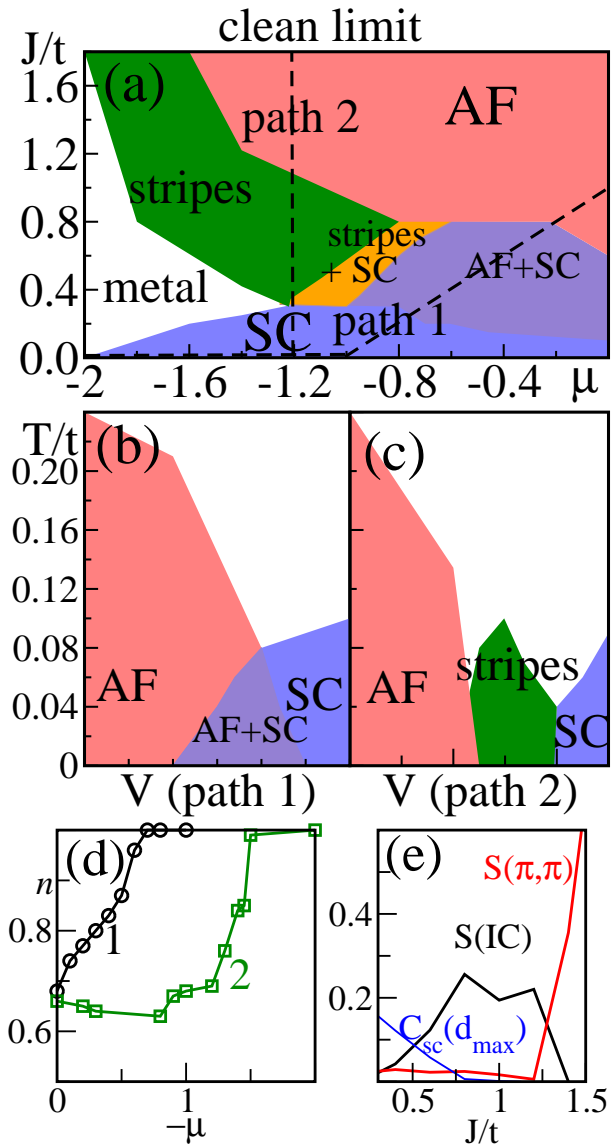

\centering{
\includegraphics[width=8cm,clip]{\mypath{prbfig1a}}
\includegraphics[width=8cm,clip]{\mypath{prbfig1b}}
}
\caption{(a) MC phase diagram for Eq.~(\ref{eq:hamfermi}) 
without disorder at low temperatures. Instead of presenting a three dimensional phase diagram we have chosen to present a
two dimensional cut along 
$V$=1-$J/2$ for simplicity. Five regions are observed: 
AF, $d$-SC, stripes, coexisting SC+AF, coexisting stripes+SC, 
and metallic. 
(b) MC phase diagram including temperature
along ``Path 1''.
(c)   MC phase diagram along ``Path 2''. 
 Lattice sizes in all cases are 8$\times$8 and 12$\times$12.
(d) $n$ vs. $\mu$ along Paths 1 and 2. Transitions along 
Path 1 appear
continuous, whereas along Path 2 there are indications of
first-order transitions. 
(e) Spin structure factor $S({\bf q})$ at ($\pi$,$\pi$) and
for incommensurate (IC) momenta.
\label{fig:3}}
\end{figure}

\noindent
where $c_{{\bf i}\sigma}$ are fermionic operators, 
$s_{\bf i}^z$=$(n_{{\bf i}\uparrow}-n_{{\bf i}\downarrow})/2$, $n_{{\bf i}\sigma}$ is
the number operator, $D$ is the lattice dimension, and
$\Delta_{{\bf i}\alpha}$=$|\Delta_{{\bf i}\alpha}|$$e^{i\phi^{\alpha}_{\bf i}}$
are complex numbers for the SC order parameter
defined at the links (${\bf i}$,${\bf i}$+$\alpha$) ($\alpha$ = unit vector along
the $x$ or $y$ directions). At $J_{\bf i}$=0, $d$-wave SC is favored close to
half-filling since
the pairing term involves nearest-neighbor sites, as in any standard mean-field approximation 
to SC.
The spin degrees of freedom (d.o.f.) are 
assumed to be Ising spins 
(denoted by $S_{\bf i}^z$). Studies with O(3) d.o.f. were found to  
lead to qualitatively
similar conclusions, but they are more CPU time consuming.
 The
parameters of relevance are $J_{\bf i}$, $\mu_{\bf i}$, and $V_{\bf i}$ ($t$
is the energy unit), that carry a site dependence to easily include
 quenched disorder which is inevitable in chemically doped compounds as the cuprates.
For a fixed configuration, 
$\{\Delta_{{\bf i}\alpha}\}$  and $\{ S_{\bf i}^z \}$,
the one-particle sector is Bogoliubov diagonalized.
In the limit $T$$\rightarrow$0, the
Bogoliubov-de Gennes equations are recovered minimizing the energy\cite{re:atkinson03,re:ghosal02,re:ichioka99}. 
Then, a standard
Monte Carlo (MC) simulation similar to those for 
Kondo-lattice models 
is carried out (details in Ref.~\onlinecite{re:dagotto02}).
One of the goals is to estimate 
$T_{\rm c}$, as well as $T^*_{\rm c}$, roughly defined as the 
temperature at which strong short-distance SC
correlations develop (more details are given below). Finally, note that model Eq.~(\ref{eq:hamfermi}) is
not derived but {\it proposed} as a possible phenomenological model for
AF vs. SC competition. The results will be shown to justify this assumption. Moreover, the
qualitative simplicity of our conclusions suggests 
that similar models will lead to similar scenarios.

\subsection{Phase Diagram in the Clean Limit}
Without quenched disorder, $V_{\bf i}$, $J_{\bf i}$ and $\mu_{\bf i}$ 
in Eq.~(\ref{eq:hamfermi}) are site independent. The standard MC analysis carried out in these 
investigations (details
provided below)
reveals that in the clean limit the 
low temperature ($T$) phase diagram, Fig.\ref{fig:3}(a), 
has a robust AF phase for electronic densities $n$$\sim$1 
and a $d$-wave SC phase for $n$$<$1. 
The $d$-wave correlation function, defined as
\begin{equation}
C^{\alpha\beta}_{sc}({\bf m})=\sum_{\bf i} \left<|\Delta_{\bf i}||\Delta_{\bf i+m}|
\cos(\phi^\alpha_{\bf i}-\phi^\beta_{\bf i+m})\right>,
\label{eq:sccorrelation}
\end{equation}
 was used to estimate
$T_{\rm c}$ as the temperature at which $d$-wave correlations 
at the largest distances for
the lattices considered here are 5\% of 
their maximum value at $|{\bf m}|$=0. The 5\% criterion is arbitrary but other criteria lead to
 identical qualitative trends, slightly shifting
the phase diagrams.
$T^*$ is deduced similarly, but using the shortest non-zero distance 
correlation function ($|{\bf m}|$=1).
The N\'eel temperature, $T_{\rm N}$, associated with the classical spins
was defined by the drastic reduction
($\leq$ 5\% of $|{\bf m}|$=0 value) of the long-distance spin order using
$C_{\rm S}({\bf m})$=$\sum_{\bf i} \left< S_{\bf i}^z S_{\bf i+m}^z \right>$, 
while $T^*_{\rm N}$ relates to
short-range spin order.
The results presented in Fig.~\ref{fig:3}(a) are not surprising since these states
are favored explicitly in Eq.~(\ref{eq:hamfermi}) by the second and fifth terms, respectively. 
However, the phase diagram presents several nontrivial interesting regions:
(i) Along ``Path 1'' in Fig.~\ref{fig:3}(a), the AF-SC transition
occurs through $local$ coexistence, 
with tetracritical 
behavior (Fig.~\ref{fig:3}(b)).\cite{re:zhang97} 
(ii) Along ``Path 2'' the AF-SC interpolating regime
has alternating doped and undoped {\it stripes} (stripes in MC data are
deduced from spin and charge structure
factors, 
and low-$T$ MC snapshots), 
and a complex phase diagram   
Fig.~\ref{fig:3}(c). These stripes evolve continuously from the  
$V$=0 limit that was studied before by Moreo et al., and as a consequence 
we refer the readers to Ref.~\onlinecite{re:moraghebi02} for further details on how stripes were identified.
It remains to be investigated if these stripes, involving SC and AF quasi-1D lines, have the
same or different origin as those widely discussed before in the high-Tc 
literature.\cite{re:tranquada97,re:poilblanc89,re:zaanen89,re:emery93,re:emery94,re:white98} At $V\neq$0, the doped regions of the stripes
have nonzero SC amplitude at the mean-field level. \cite{re:chen03} In view of the dramatically different behavior along
Paths 1 and 2, we conclude that
in our model {\it there is no unique AF$\rightarrow$SC path}.
This is in agreement with experiments since $\rm La_{2-{\it x}} Sr_{\it x} Cu O_4$ (LSCO) and others have
stripes,\cite{re:tranquada95,re:emery90,re:zhou01}
while $\rm Ca_{2-{\it x}} Na_{\it x} Cu O_2 Cl_2$ 
has a more complex inhomogeneous 
pattern.\cite{re:kohsaka04} Both, however, become
SC with increasing $x$. This suggests
that {\it the underdoped region of Cu-oxides
may not be universal.}

\begin{figure}
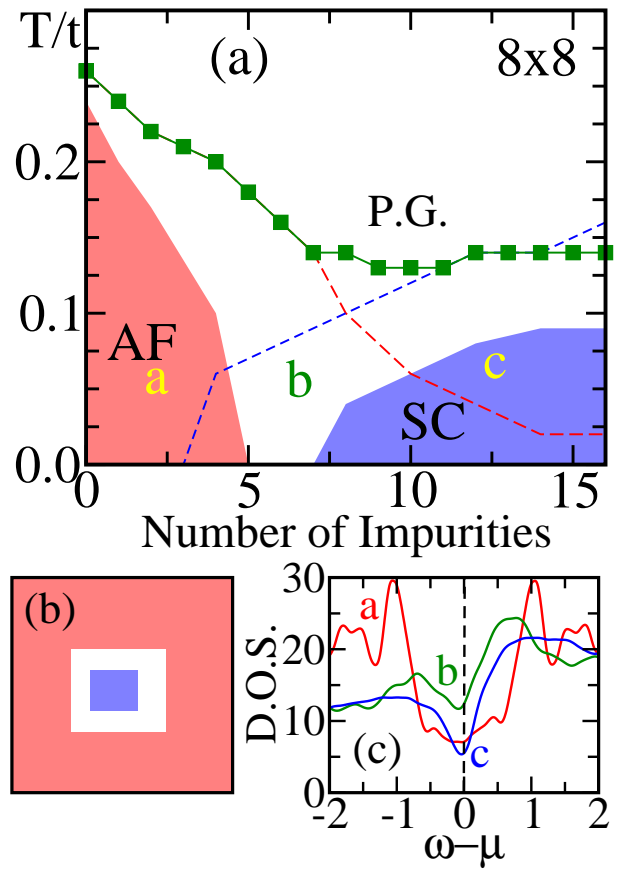

\centering{
\includegraphics[width=8cm,clip]{\mypath{prbfig2a}}
\includegraphics[width=8cm,clip]{\mypath{prbfig2b}}
}
\caption{
(a) Plaquette impurity schematic representation.
Disorder may have several forms, but here we mimic Sr-doping in 
single-layers. Sr$^{2+}$ replaces La$^{3+}$, 
above the center of a Cu-plaquette in the Cu-oxide square lattice. 
Then, as hole carriers are added, a hole-attractive 
plaquette-centered potential should also be incorporated. Near the center 
of this potential, $n$ should be sufficiently 
reduced from 1 that, phenomenologically, tendencies to SC should be expected.
To interpolate between the SC central plaquette and the AF background,
a plaquette `halo' with no dominant tendency was introduced
Parameters are chosen such that the blue (black) region favors 
superconductivity, $(J,V,\mu)$=$(0.1,1.0,-1.0)$, with a surrounding 
white region where $(J,V,\mu)$=$(0.1,0.1,-0.5)$ with no order prevailing.
The impurity is embedded in a background (red, dark gray) 
that favors the AF state, $(J,V,\mu)$=$(1.0,0.1,0.0)$.
However, the overall conclusions found here are simple,
and independent of the disorder details.
(b) MC phase diagram for model Eq.~(\ref{eq:hamfermi}) 
including quenched disorder (lattices studies are 
8$\times$8 (results shown) and 12$\times$12). 
Shown are $T_{\rm c}$ and $T_{\rm N}$ vs. number
of impurities (equal to number of holes). 
The SC and AF regions with short-range 
order (dashed lines), and 
$T^*$ as obtained from the PG (dot-dashed line) are also indicated. 
(c) DOS at points $a$, $b$, and $c$ of (a), with a PG.
\label{fig:4}}
\end{figure}

\begin{figure}
\centering{
\includegraphics[width=8cm,clip]{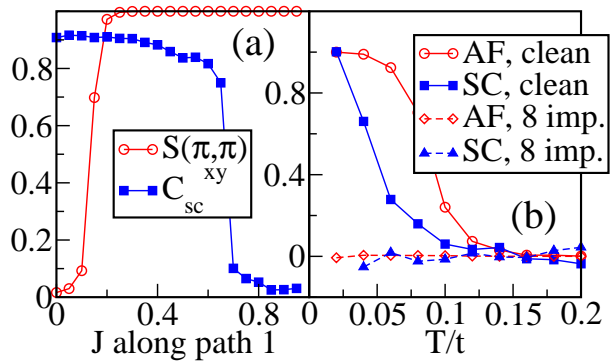}
}
\caption{\label{fig:5}(a) MC $S(\pi,\pi)$ and superconducting
correlation at maximum distance, $C_{sc}^{xy}$, Eq.~(\ref{eq:sccorrelation}), 
 along ``Path 1'' (Fig.~1(a)),
using an $8 \times 8$ lattice and at low temperature $T$=0.02. 
(b) AF and SC correlations at maximum distance for the model with 8 impurities and
without disorder (clean). The latter is  the point $J$=$0.6$ of ``Path 1''. 
The disordered case corresponds to 8 plaquettes on a 12$\times$12 lattice.
Typically, 5,000 sweeps were used for thermalization and for measurements.
With quenched disorder, many points were obtained after averaging over
$N_{\rm dis}$=10 disorder realizations. 
The results were found to mildly depend on the disorder configuration, 
thus many results were obtained with a smaller $N_{\rm dis}$.
} 
\end{figure}

\subsection{Phase Diagram with Quenched Disorder}

Our results become even more interesting upon introducing quenched disorder,
with a MC phase diagram shown in Fig.~\ref{fig:4}(b). The
similarity with the widely accepted phase diagram of the cuprates is clear.
The disorder has opened a hole-density ``window'' 
where none of the two competing orders dominates. 
Inspecting ``by eye'' the dominant MC configurations (snapshots) at
low-$T$ in this intermediate regime reveals a patchy system with 
slowly evolving islands of SC
or AF, and random orientations of the local order parameters, leading to an
overall disordered ``clustered'' state. 
In Fig.~\ref{fig:4}(b), a new temperature scale $T^*$  at which the fermionic 
density-of-states (DOS) develops a {\it pseudogap} (PG) 
(Fig.~\ref{fig:4}(c)) was also unveiled. 
The AF and $d$-SC regions both have a gap (smeared
by $T$ and disorder, but nevertheless with 
recognizable features). But even the ``disorder'' regime (case b in
Fig.~\ref{fig:4}(c)) has a
PG. MC snapshots explains this behavior: in the
disordered state there are small SC or AF regions, as
explained above. Locally each has a smeared-gap DOS, either AF or SC. Not
surprisingly, the mixture presents a PG. 
The behavior of $T^*$ vs.
 $x$ is remarkably similar to that found experimentally. 
{\it The cuprates' PG may arise from
an overall-disordered clustered state with local {\rm AF} or {\rm SC}
tendencies},
without the need to invoke other exotic states. This PG is correlated 
with robust short-range correlations (dashed
lines in Fig.\ref{fig:4}(b), see caption for details.). 

The numerical procedure that led to Figs.~\ref{fig:3} and \ref{fig:4} 
is standard, well-known
in the manganite context where formally similar models are widely 
studied,\cite{re:dagotto02,re:dagotto01,re:tokura00b} thus here only a few representative 
results will be presented for completeness. For instance, 
Fig.~\ref{fig:5}(a) shows the order parameters along ``Path 1'' 
of Fig.~\ref{fig:3}(a), indicating how
each ordered region was determined in the clean limit. 
Clearly, a region of coexistence can
be observed. Similar data were used to complete the phase diagram. 
Likewise, Fig.~\ref{fig:5}(b) contains 
the order parameter vs. $T$ with and without disorder, 
in the interesting region of couplings and doping.
There is a drastic difference between clean and dirty limits,
the latter showing no global dominant order. However, examining
relevant MC configurations, small SC and AF clusters with
random orientations are found.

\section{Model II: Landau Ginzburg}
The results reported thus far,
based on Eq.~(\ref{eq:hamfermi}), have already revealed interesting
information, namely the possible paths from AF to SC, and a proposed explanation of the glassy
state as arising from the inevitable quenched disorder in the samples. However, the inhomogeneous nature
of the clustered region suggests that percolative phenomena may be at
work, and larger clusters are needed. To handle
this issue, 
another model containing \emph{only} classical d.o.f. is proposed,
with low-powers interactions typical of Landau-Ginzburg (LG) approaches:
\begin{eqnarray}\nonumber
&H&=r_1 \sum_{\bf i}|\Delta_{\bf i}|^2 +
\frac{u_1}{2}\sum_{\bf i}|\Delta_{\bf i}|^4 
+ \sum_{{\bf i},\alpha}\rho_2({\bf i},\alpha){\bf S}_{\bf i}\cdot {\bf S}_{{{\bf i}}+\alpha}\\  \nonumber
&-&\sum_{{\bf i},\alpha} \rho_1({\bf i},\alpha)|\Delta_{\bf i}||\Delta_{{\bf i}+\alpha}|
\cos(\Psi_{\bf i}-\Psi_{{\bf i}+\alpha})+ {\it r_2} \sum_{\bf i} |{\bf S}_{\bf i}|^2 \\ 
&+&\frac{u_2}{2} \sum_{\bf i} |{\bf S}_{\bf i}|^4 +  u_{12}\sum_{\bf i}
|\Delta_{\bf i}|^2|{\bf S}_{\bf i}|^2.
\label{eq:hamgl}
\end{eqnarray} 

\noindent
The fields $\Delta_{\bf i}$=$|\Delta_{\bf i}|{\rm e}^{i\Psi_{\bf i}}$ 
are complex numbers representing the SC order parameter. The
classical spin at site ${\bf i}$ is 
${\bf S}_{\bf i}$=$|{\bf S}_{\bf i}|(\sin(\theta_{\bf i})\cos(\phi_{\bf i}),
\sin(\theta_{\bf i})\sin(\phi_{\bf i}),\cos(\theta_{\bf i}))$.
$\rho_1({\bf i},\alpha)$=$1 - \rho_2({\bf i},\alpha)$ 
is used as the analog of $V$=1-$J/2$ of the previous model to reduce the multiparameter character of the investigation,
allowing an AF-SC interpolation changing just one parameter.
$\alpha$ denotes the two directions $\hat{x}$ and $\hat{y}$ in 2d,
 and also $\hat{z}$ for multilayers.
$\rho_2({\bf i},\alpha)$ was chosen to be isotropic, i.e., 
$\alpha$-independent. 

\subsection{Basic Properties}
Clearly, the lowest-energy state for $\rho_2$=0 is a homogeneous SC state
%
(if $\rho_1({\bf i},\alpha)$=$\rho_1^0$$>$0). 
%
When $\rho_1$=0 the
lowest-energy state is AF
(if $\rho_2({\bf i},\alpha)$=$\rho_2^0$$>$0). In the clean limit, 
this model was
already studied in the SO(5) context, where the reader is referred
for further details. Our approach without disorder has similarities
with SO(5) ideas\cite{re:zhang97} where the AF/SC
competition as the cause of the high-$T_{\rm c}$ phase diagram
was extensively discussed although nowhere in our investigations we need to invoke a higher symmetry group.
The relevance of tetracriticality 
in $\rm La_2 Cu O_{\it 4+\delta}$ has also
been remarked by E. Demler {\it et al.}\cite{re:demler01}
and Y. Sidis.\cite{re:sidis01} In the present work, 
disorder is introduced by adding a randomly selected bimodal contribution, i.e.
$\rho_2({\bf i},\alpha)$=$\rho_2^0$$\pm$$W$,
where $W$ is the disorder strength ($W$=0 is the clean limit).
It is expected that other forms of disorder will lead to similar
results.

\subsection{Phase Diagram}
\begin{figure}
\centering{
\includegraphics[clip,width=7cm]{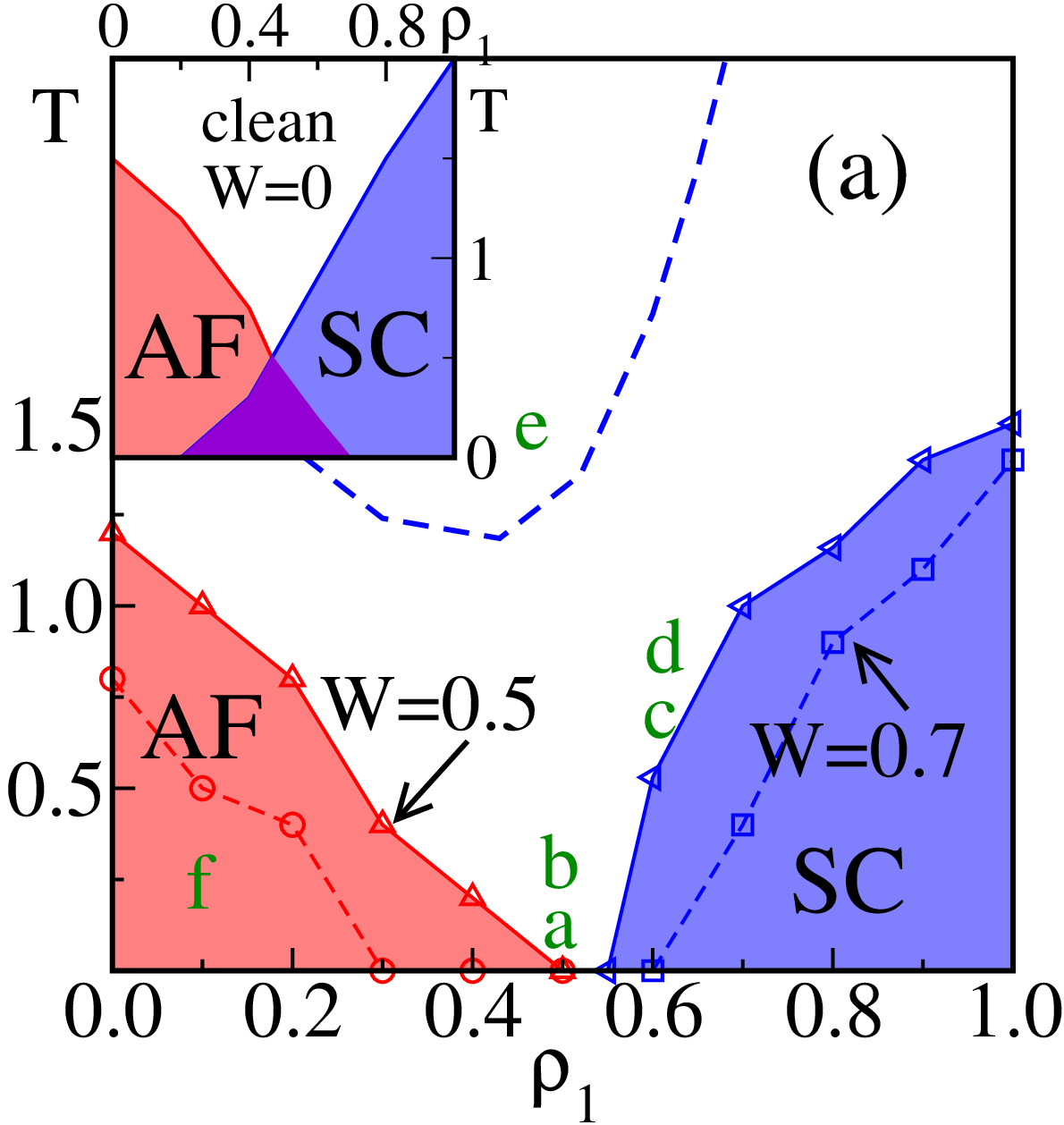}}
\parbox{8.2cm}{
\parbox{4cm}{\includegraphics[clip,width=4cm]{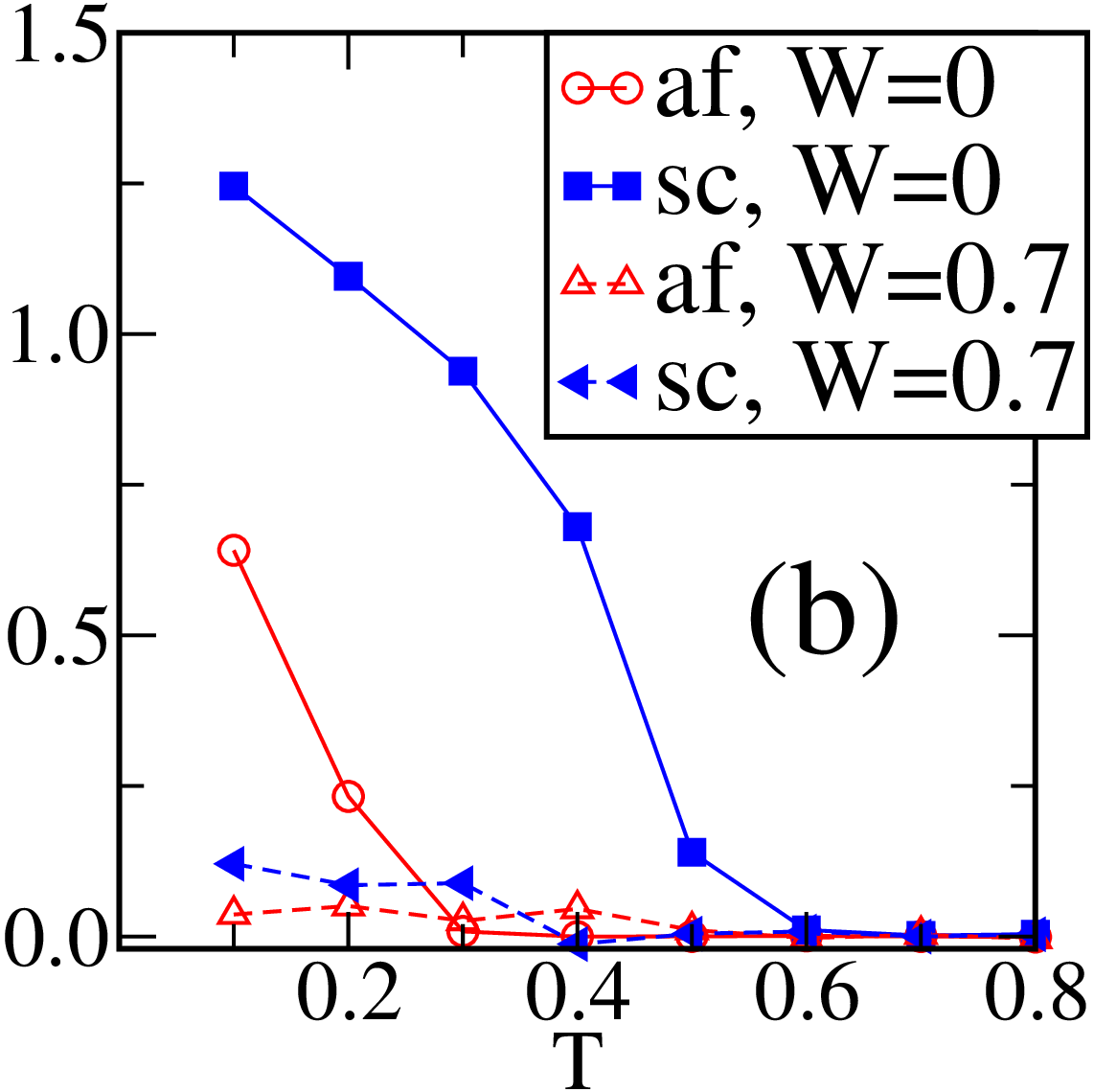}}
\parbox{4cm}{\includegraphics[clip,width=4cm]{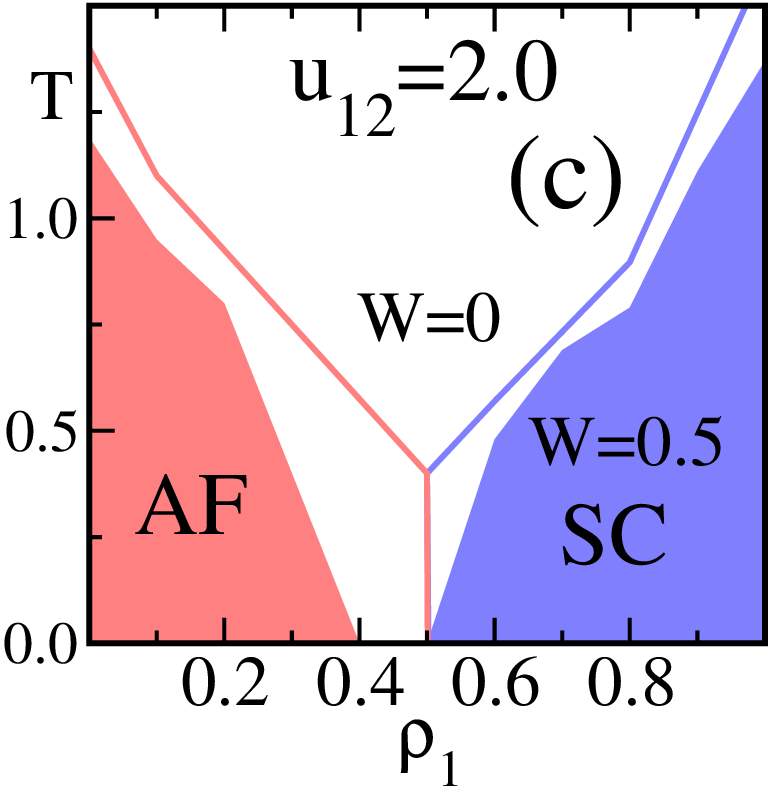}}
}
\caption{(a) MC phase diagram (for Eq.~(\ref{eq:hamgl})) at
$u_{12}$=$0.7$.
Parameters are $r_1$=-1,
$r_2$=$-0.85$, and $u_1$=$u_2$=1 but the
conclusions are not dependent on coupling fine-tuning.
Spin
$C_{\rm spin}({\bf m})$=$\frac1N\sum_{{\bf i}}\left<
  {\bf S}_{{\bf i}}\cdot{\bf S}_{{\bf i}+{\bf m}}\right>$ and
SC correlations
$C_{\rm SC}({\bf m})$=$\frac1N\sum_{{\bf i}}
|\Delta_{{\bf i}}||\Delta_{{\bf i}+{\bf m}}|
\cos(\Psi_{{\bf i}}-\Psi_{{\bf i}+{\bf m}})$ were measured.
The behavior of these functions at the largest (shortest) distance
determine $T_{\rm c}$ and $T_{\rm N}$ ($T^*$) (same criteria
as for Eq.~(\ref{eq:hamfermi})).
With disorder,
the phase diagram (shown) has an intermediate ``clustered'' state
with short-range order. $T^*$ is
also indicated (dashed line).
Note the similarity with
Fig.~\ref{fig:4}(b).
{\it Inset:} results at $W$=0 showing
tetracriticality (magenta (dark)
indicates SC-AF coexistence).
(b) AF and SC correlations at maximum distance for the model
Eq.~(\ref{eq:hamgl}) without and with
disorder ($W$=$0.0$ and 0.7, respectively).
$\rho_1$=$0.5$ and $u_{12}$=$0.7$ were used, using a 24$\times$24
lattice.
Typically, for the LG model 25,000 sweeps were used for thermalization
and measurements.
(c) MC phase diagram of
model Eq.~(\ref{eq:hamgl}) at $u_{12}$=$2$.
The clean case ($W$=$0$, solid lines) is bicritical-like,
but with disorder $W$=$0.5$
a clustered region between SC and AF opens as well.
\label{fig:1}}
\end{figure}

Monte Carlo results for Eq.~(\ref{eq:hamgl}) are in Fig.~\ref{fig:1}a, for 
 ``weak''  
coupling $u_{12}$=0.7, which leads to tetracritical 
behavior. Both at $W$=0 and $W$$\neq$0, 
the qualitative similarity with fermionic model results
(Figs.\ref{fig:3}(b) and \ref{fig:4}(b)) is clear. Coexisting SC and AF clusters
appear in MC snapshots (not shown). 
Then, both models share a similar phenomenology, and Eq.~(\ref{eq:hamgl}) can
be studied on larger lattices. The only important difference between the two models is that
  Eq.~(\ref{eq:hamgl}) cannot lead to doped-undoped
stripes, but the more general case Eq.~(\ref{eq:hamfermi}) does. 
Fig.~\ref{fig:1}(b) illustrates how the phase
diagram, Fig.~\ref{fig:1}(a), was obtained. 
For completeness, note that increasing the coupling $u_{12}$ a first-order SC-AF transition can be obtained. However, the
addition of disorder leads to a very similar phase diagram as in the case of $u_{12}=0.7$.
This is shown in Fig.~\ref{fig:1}c and is the equivalent of
Fig.~\ref{fig:1}a in the regime of ``strong'' coupling.

Some of the experimental predictions related with our 
SC-AF clustered state are simple (the most elaborated ones are in the next section). 
In most ways a very underdoped cuprate
can be tested, there should be two components in the data. 
For instance, a typical photoemission spectra in our framework should have 
two clearly distinct coexisting signals.
This result, which will be discussed in more detail in a future publication, is compatible with photoemission
experiments for $x$=0.03 LSCO, that reveal spectral 
weight in the node direction of the $d$-wave superconductor 
even in the insulating glassy regime.\cite{re:yoshida03} Nodal $d$-wave SC particles surviving to low $x$
was observed in Ref.~\onlinecite{re:hosseini03}.

\section{Colossal Effects in Cuprates}

\begin{figure}
\parbox{8.2cm}{\parbox{4.2cm}{\includegraphics[clip,width=4.2cm]{\mypath{prbfig5left}}}
\parbox{3.8cm}{\includegraphics[clip,width=3.6cm]{\mypath{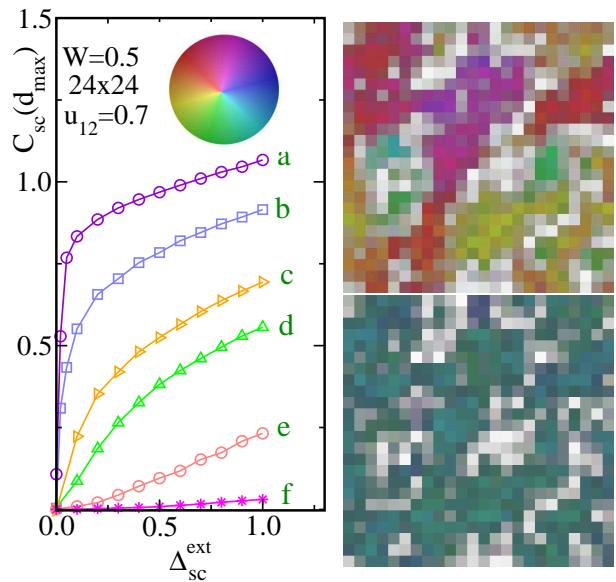}}
\includegraphics[clip,width=3.6cm]{\mypath{d5sc}}}}
\caption{{\it Left:} $C^{\rm max}_{\rm SC}$ vs. $\Delta^{\rm ext}_{\rm SC}$ (see text)
on a 24$\times$24 lattice, with $u_{12}$=$0.7$ 
and $W$=$0.5$, at the five points {\it a-f} indicated in Fig.~\ref{fig:1}a.
A ``colossal'' effect is observed in $a$ and $b$ 
where the $\Delta^{\rm ext}_{\rm SC}$=0 state is ``clustered''. 
A much milder (linear)
effect occurs far from the SC phase ($e$ and $f$).
MC snapshots are shown at $\Delta^{\rm ext}_{\rm SC}$=$0.0$ {\it right, top}
and $\Delta^{\rm ext}_{\rm SC}$=$0.2$ {\it right, bottom}, 
both at $T$=$0.1$ and $\rho_2$=$0.5$, 
using the same quenched-disorder configuration. 
The color convention is explained in
the circle (colors indicate the SC phase, while 
intensities are proportional to
$Re(\Delta_{\bf i})$). The AF order parameter is not shown. 
The multiple-color nature of the upper snapshot, reflects
a SC phase randomly distributed (i.e. an overall non-SC state). 
However, a small external field rapidly aligns those phases, 
leading to a coherent state ({\it bottom}). 
}
\label{fig:2}
\end{figure}

One of the main results of these investigations is that
 the models studied here can
present ``colossal'' effects, very similarly in spirit as it occurs in manganites. 
Consider a typical clustered state (Fig.~\ref{fig:2}(b)) 
found by MC simulations in the disordered region.
This state has preformed local SC correlations -- nanoscale
regions having robust SC amplitudes within each region, but no SC phase
coherence between different regions -- rendering the state 
globally non-SC (the averaged correlation at the
largest distances available, $C^{\rm max}_{\rm SC}$, 
is nearly vanishing). Let us now introduce an
artificial SC ``external field'', which can be imagined as caused by the 
proximity of a layer with robust SC order (e.g., 
comprised of a higher-$T_{\rm c}$ material).
In practice, this is achieved in the calculations
by introducing a term
$|\Delta^{\rm ext}_{\rm SC}|\sum_{\bf i}\rho_1({\bf i},\hat{z})|\Delta_{{\bf i}}|
\cos(\Psi_{{\bf i}})$, where $\Delta^{\rm ext}_{\rm SC}$ acts as an external
field for SC. The dependence of $C^{\rm max}_{\rm SC}$
with $\Delta^{\rm ext}_{\rm SC}$ is simply remarkable (Fig.~\ref{fig:2}(a)). While at
points {\it e} and {\it f}, located far from the SC region in Fig.~\ref{fig:1}a, 
the dependence is the expected one for a featureless state (linear), 
the behavior closer to SC and small temperatures
is highly nonlinear and unexpected. For example, at point {\it a}, 
$C^{\rm max}_{\rm SC}$ vs. $\Delta^{\rm ext}_{\rm SC}$ has 
a slope (at $\Delta^{\rm ext}_{\rm SC}$=0.02) which is
{\it $\sim$250 times larger} than at {\it e} ($\sim$13 times larger than
at $W$$=$0, same $T$, $\rho_2$, and $u_{12}$.).

The reason for this
anomalous behavior is the clustered nature of the states. This is 
shown in the state Fig.~\ref{fig:2}(c), contrasted with (b), where
a relatively small field -- in the natural units of the model --
nevertheless led to a quick alignment of SC phases, producing a globally SC
state, as can be inferred from the uniform color of the picture.
 \emph{Having preformed SC puddles vastly increases the SC susceptibility.} 
Since Fig.~\ref{fig:2}(a) was obtained in a trilayer geometry
it is tempting to speculate 
that  the proximity
of SC layers to a non-SC but clustered state, can naturally lead to 
a GPE over long distances, as observed experimentally
in a similar geometry.\cite{re:bozovic04,re:decca00,re:quintanilla03}

\begin{figure}
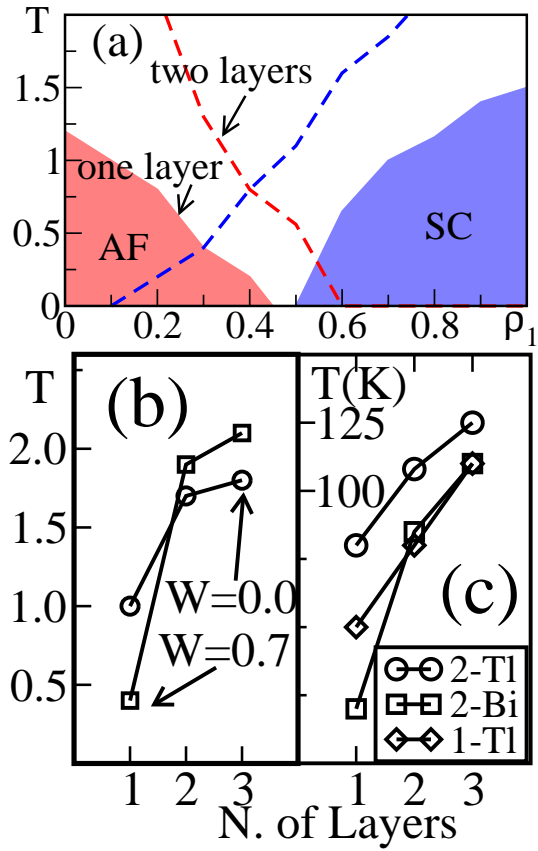

\centering{
\includegraphics[clip,width=7.0cm]{forprbfig7b}
\includegraphics[clip,width=7.0cm]{forprbfig7a}
}
\caption{\label{fig:layers}(a) MC phase diagram (for Eq.(\ref{eq:hamgl})) at
$u_{12}$=$0.7$.
Parameters are $r_1$=-1,
$r_2$=$-0.85$, $u_1$=$u_2$=1, $W=0.5$ with one layer (solid colors) and two layers (dashed line). 
The addition of an extra layer increases the critical temperature of the superconductor as well as the
N\'eel temperature.
(b) $T_{\rm c}$ vs. $N_{\rm \ell}$ for $u_{12}$=0.7,  
$\rho_2$=$0.3$, $W$=0.7, and 24$^2$$\times$$N_l$ clusters. Shown are
results with and without disorder.
(c) The experimental $T_{\rm c}$ (in K) is shown 
for three HTS families, as indicated, up to 3 layers (data from Ref.~\onlinecite{re:burns92}).
}
\end{figure}

\section{Dependence of $T_{\rm c}$ with the number of layers}
The nanoscale clusters 
also leads to a proposal for explaining
the {\it rapid }
increase of $T_{\rm c}$ with the number of Cu-oxide layers $N_{\rm \ell}$,
found experimentally, at least up to 3 layers.
In this effort, the MC phase diagrams of single-, bi-, and tri-layer systems
described by Eq.~(\ref{eq:hamgl}), with and without disorder, were calculated
using exactly the same parameters 
(besides a coupling $\rho_2({\bf i},\hat{z})$, equal to those along $\hat x$ 
and $\hat y$, to connect the layers). It was  
clearly observed that {\it the single layer
has a substantially lower $T_{\rm c}$ than the bilayer}. 
This can be understood in
part from the obvious critical fluctuations that are stronger in 2D than 3D.
But even more important, cluster percolation
at $W$$\neq$0 is more difficult
in 2D than 3D (since otherwise 2D disconnected clusters may become linked
through an interpolating cluster in the adjacent layer). 
Then, in the phenomenological approach presented here 
it is natural that $T_{\rm c}$ 
increases fast with $N_{\rm \ell}$, when changing from 1 to 2 layers as shown in
Fig.~\ref{fig:layers}a.
This concept is even {\it quantitative} -- up to a scale -- 
considering the similar shape of $T_{\rm c}$ vs. $N_{\rm \ell}$
found both in the MC simulation and in experiments 
(see Figs.\ref{fig:layers}b-c. Note that the subsequent decrease of $T_{\rm c}$
for 4 or more layers observed experimentally 
could be caused by inhomogeneous doping, 
beyond our model). \emph{Our MC results suggest that the large
variations of $T_{\rm c}$'s known to occur in single-layer cuprates can be
attributed to the sensitivity of 2D systems to disorder.}
As $N_{\rm \ell}$ increases (the system becomes more 3D),
the influence of disorder {\it decreases}, both in experiments\cite{re:eisaki03}
and simulations.

\section{Conclusions}

\begin{figure}
\includegraphics[width=7.5cm]{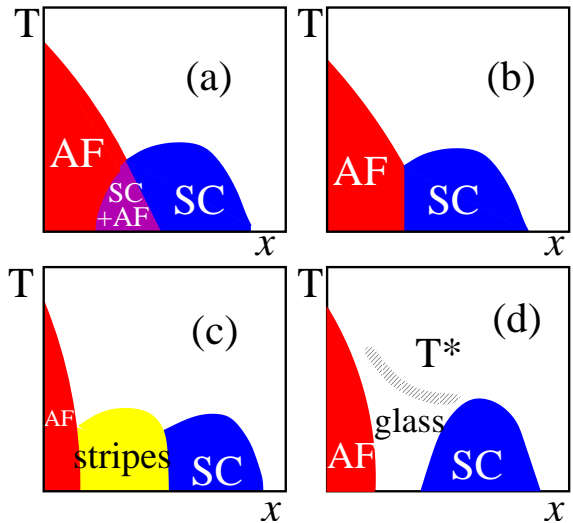}
\caption{\label{fig:sketchphases}  Schematic representation of the phase diagrams that our models
 show in the clean (a,b,c) and dirty (d) limits. The theory discussed
 in this paper shows the possible appearance of regions with {\it local}
 coexistence of AF and SC (panel a), or a first-order transition
 separating AF from SC (panel b) with the first-order character of
 the transition possibly continuing in the AF-disordered and SC-disordered
 transitions, or an intermediate striped regime (panel c). Possibilities
 (a) and (b) have already been discussed in Ref.~\onlinecite{re:zhang97}, although here we do
  not invoke a higher symmetry group such as SO(5). The main result 
  contain in this figure is the proposed phase diagram in the presence
  of quenched disorder (panel d). Shown are the glassy region, proposed
  to be a mixture of SC and AF clusters, and the $T^*$ where local
  order starts upon cooling. This phase diagram has similarities with
 those proposed before for manganites,\cite{re:dagotto02,re:dagotto01}
 and certainly it is in excellent agreement with the experimental phase diagram of LSCO}.
\end{figure}

\begin{figure}
\includegraphics[width=4cm]{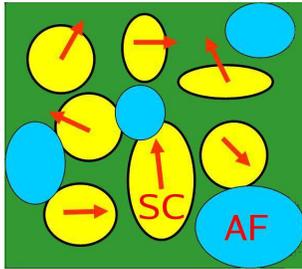}
\caption{\label{fig:clustered}  Schematic representation of the ``glassy'' state that separates the
 SC and AF regions. The arrow indicates the phase of the SC order 
 parameter.}
\end{figure}

Summarizing, here simple phenomenological models for phase 
competition showed that -- depending on details --
different cuprates could have stripes, local coexistence, first-order
transitions, or a glassy clustered state interpolating 
between AF and SC phases. Figure \ref{fig:sketchphases} illustrates our proposed possibilities.
In Cu-oxides where the glass state is realized, namely where SC puddles are
present,  this study revealed the possibility of
colossal effects. A schematic representation of the proposed glassy state with colossal effects is in Fig.~\ref{fig:clustered}.
This proposal could provide rationalization of recent results in trilayer thin-film
geometries.\cite{re:bozovic04,re:decca00,re:quintanilla03}

After submission of this work, we learned of interesting experimental
efforts that complement the discussion presented here: 
{\it (1)} In
Ref.~\onlinecite{re:asulin04}, further evidence of an anomalous proximity effect in the
cuprates is presented. These results add to those of 
Ref.~\onlinecite{re:bozovic04,re:decca00,re:quintanilla03}, 
showing
that the anomalous effects are real.
{\it (2)} In Ref.~\onlinecite{re:sanna04,re:sanna04b}, 
the phase diagram of YBCO was recently
investigated in the presence of Ca doping. Among many results,
it was shown that a glassy state is generated between the AF and SC
states in Ca-doped YBCO,
with a phase diagram very similar to that in LSCO and our Fig.~\ref{fig:sketchphases}(d). 
This result
suggests that {\it Ca-undoped} YBCO may have either a region of local 
coexistence of SC and AF or a first-order transition separating them
 (as in Fig.~\ref{fig:sketchphases}(a,b)),
and only with the help of extra quenched disorder is that a glassy
state is generated. Then, the generic phase diagram of the cuprates --
which usually is considered to be that of LSCO -- may not be as universal
as previously believed, as discussed in this publication.
Our study showing that bilayered systems are
more stable than single layers with respect to disorder is also compatible
with the experimental results of Ref.~\onlinecite{re:sanna04,re:sanna04b}, 
namely the 1-layer material is
more likely to have a glassy state between AF and SC than 2- or higher
layer materials.
{\it (3)} Our effort has already induced interesting theoretical work\cite{re:arrachea04}
in the context of $J$-$U$ models.
{\it (4)} Theoretical work\cite{re:tu04} closely related to our proposed glassy state in 
Fig.~\ref{fig:sketchphases} has addressed inhomogeneous Josephson phases
near the superconductor-insulator transition.
{\it (5) } Recent neutron and Raman scattering investigations applied to
La$_2$CuO$_{4.05}$  has shown the coexistence of SC and AF phases
in this compound.\cite{re:gnezdilov04}
{\it (6)} Finally, our results have similarities with those
recently discussed in the context of {\it Bose metals} as well.\cite{re:dalidovich03}

The study also provided predictions for photoemission experiments (to be discussed elsewhere) and
a simple explanation for the $T_{\rm c}$ increase
with $N_{\rm \ell}$  (another explanation can be found in
 Ref.~\onlinecite{re:chakravarty04}).
Clustered states are crucial in manganites and other compounds,\cite{re:lance02}
and this analysis predicts its potential
relevance in HTS materials as well.

\begin{acknowledgments} 
Work supported by NSF grants DMR 0122523 and 0312333.
Conversations with S. L. Cooper, J. Tranquada, S. Sachdev, M. Greven,
S. Chakravarty, and S.C. Zhang are gratefully acknowledged.
\end{acknowledgments}

\bibliography{thesis}

\end{document}